# Accurate Single-Ended Measurement of Propagation Delay in Fiber Using Correlation Optical Time Domain Reflectometry

Florian Azendorf, *Member, OSA*, Annika Dochhan, *Member, IEEE,* and Michael H. Eiselt, *Fellow, OSA, Senior Member, IEEE*

*Abstract*—A correlation optical time-domain reflectometry (C-OTDR) method is presented, which measures the propagation delay with an accuracy of a few picoseconds. This accuracy is achieved using a test signal data rate of 10 Gbit/s and employing cross-correlation and pulse fitting techniques. In this paper we introduce and evaluate the basic signal processing steps, investigate the measurement accuracy, and discuss applications for monitoring link delay and chromatic dispersion of long fiber spans as well as temperature sensing applications.

## I. Introduction

ASYMMETRIC delay in optical transmission systems has received increased attention during the recent years, as synchronization applications using the IEEE 1588v2, or precision time protocol (PTP), standard rely on the same propagation latency in both directions between the master clock and the slave clock. Any asymmetry in the propagation delay leads to an error in the slave clock synchronization. Symmetry requirements are on the order of a few nanoseconds. Likewise, the common public radio interface (CPRI) [1], used for mobile front haul communication between the base band unit and the remote radio head, allows for a maximum asymmetry of 8 ns between the two propagation directions. An even stricter symmetry requirement is posed by the application of analog radio over fiber feeding a phase array antenna [2]. Here the delay differences between the signals destined for the individual antenna elements must not differ by more than 30 degrees of the RF carrier, which corresponds to a delay difference of only 3.2 ps for a 26-GHz radio signal.

For all these applications, asymmetric propagation delay can be corrected for, if it can be determined with sufficient accuracy. Using an optical time domain reflectometer (OTDR), the propagation delay of a fiber can be measured from a single end by transmitting an optical pulse into the fiber and recording the backscattered and reflected signals over time. However, to obtain a sufficient signal-to-noise ratio, typically pulses with a width of several nanoseconds are used, leading to a measurement accuracy in the same order of magnitude. To improve the accuracy, correlation techniques or spread spectrum techniques [3,4] were introduced a few years after the first OTDRs were proposed [5,6]. More recently, other techniques were presented, like the chaotic correlation technique [7], where the probing sequence is generated from a chaotic light source. This leads to an unambiguous probe signal per measurement and avoids the need for external modulation but requires an additional optical receiver to record the probe signal and a more complex signal processing, as well as amplification of the chaotic light by an optical amplifier.

In this paper, we report on a modification of the standard correlation technique OTDR (C-OTDR), by using a pulse sequence at a rate of up to 10 Gbit/s. In addition to the higher data rate, the backscattered signal is over-sampled five times and, after cross-correlation with the probe sequence, a Gaussian or raised-cosine pulse shape is fitted to the correlation peak. This further improves the timing resolution of the round-trip propagation delay to a few picoseconds even over a fiber length of 100 km. This is, to our knowledge, the best C-OTDR accuracy achieved so far. In addition to demonstrating the method in a laboratory setup, we report on the integration of the setup in a field-programmable gate array (FPGA) with a commercially available pluggable transceiver (SFP+) as the optical front end [8]. Using standard telecommunication components and avoiding the need for optical amplification, the implementation cost is strongly reduced, enabling network-wide deployment to monitor the dynamics of fiber propagation delay.

In section II, the measurement principle of this C-OTDR is introduced, including the optical setup and the signal processing steps. In section III, the accuracy of the measurement technique is determined and discussed. In section IV, the use of the C-OTDR for monitoring the delay of an optical link is discussed with a demonstration of the measurement for a 100-km single mode fiber span. In section V, the C-OTDR technique is applied

This work has received funding from the European Union´s Horizon 2020 research and innovation programme under grant agreement No 762055 (BlueSpace project) and from the German ministry for education and research (BMBF) under grant No 16KIS0989K (OptiCON project).

F. Azendorf, A. Dochhan, and M. H. Eiselt are with ADVA, Maerzenquelle 1-3, 98617 Meiningen, Germany (e-mail: {fazendorf, adochhan, meiselt}@adva.com).

to the investigation of the temperature behavior of the propagation delay for short fiber jumpers as well as for fibers in a buried cable. Furthermore, we show in section VI that it is feasible to measure the chromatic dispersion of a fiber under test from one side. In section VII, an outlook for new sensing applications using the C-OTDR technique is discussed. Finally, in section VIII we discuss the integration of the C-OTDR using standard telecommunication components.

## II. MEASUREMENT SETUP

### A. Optical Setup

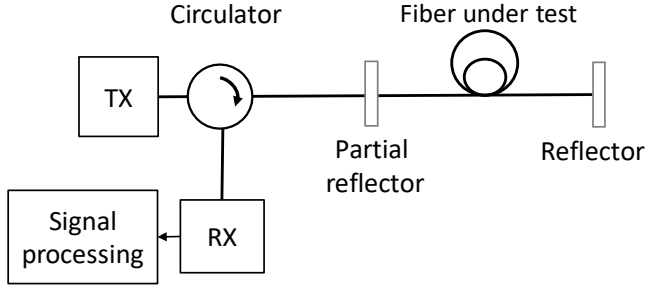

Fig. 1. Schematic Correlation-OTDR setup.

The schematic C-OTDR setup is shown in Fig. 1. A bit sequence, followed by a series of zeros, is transmitted via a circulator into the fiber under test. We used a Mach-Zehnder modulator to modulate a 10-Gbit/s signal onto a continuous wave signal from a tunable laser. Part of the signal is reflected back at a partial reflector which has a reflection coefficient of 1 to 10 percent located at the fiber input. The remaining part of the signal travels further along the fiber, is partially reflected at reflection points in the fiber and at the fiber end, where a highly reflective device can be used to maximize the reflected power. The reflected signals are directly detected on a photodiode / transimpedance amplifier combination and recorded on a real-time oscilloscope, where a bandwidth of 10 GHz and a sampling rate of 50 GS/s were used. The overall transmitted sequence length, including the trailing zeros, must be longer than the round-trip time through the fiber under test. The sequence is repeated, and the oscilloscope averages the received signals, triggered by a start-of-transmission signal from the transmitter, such that signals reflected at the same location in the fiber are always recorded at the same time on the oscilloscope. The averaged signal is then cross-correlated with the transmitted bit sequence (excluding the trailing zeros) and the round-trip time to the reflection points is determined, as described below.

### B. Signal Processing

A precise location of the reflection point can be obtained by cross-correlation of the reflected received signal with the transmitted sequence. The time evolution of the transmitted signal, carrying the (uni-polar, 0/1) data sequence $d_{TX}(n)$, can be described as

$$s(t) = \sum_{n=0}^{N-1} d_{TX}(n) \cdot p(t - nT). \quad (1)$$

Here, $N$ is the length of the sequence, $T$ is the bit duration, and $p(\tau)$ is the pulse shape, centered at τ=0.
Outside the sequence, only "zero" bits are transmitted:

$$d_{TX}(n) = 0 \text{ for } n \notin [0, \dots, N-1]. \quad (2)$$

The sequence $d_{TX}(n)$ is chosen such that the aperiodic auto-correlation function of the corresponding bi-polar sequence

$$d_{BP}(n) = 2 \cdot d_{TX}(n) - 1 \quad (3)$$

is non-zero only for a shift of zero:

$$\sum_{n=0}^{N-1} d_{BP}(n) \cdot d_{BP}(n+m) = N \cdot \delta(m). \quad (4)$$

Here, δ($n$) is the Kronecker delta function. While usually (4) is not met for a single sequence, two complementary Golay sequences [10] are used below, whose correlation sum meets the requirements of (4).

We also require the sequence to be bias-free, such that

$$\sum_{n=0}^{N-1} d_{BP}(n) = 0. \quad (5)$$

While this is not strictly true for both complementary Golay sequences, the resulting error is small.

The signal is reflected and received, after the propagation time $t_p$. The received signal is down sampled to the bit rate with an initial sampling time $t_0$ and a sample interval of $T$. The samples $s(k)$ can be described as

$$s(k) = r \sum_{n=0}^{N-1} d_{TX}(n) \cdot p(t_0 + kT - t_p - nT). \quad (6)$$

Here, $r$ is a factor taking into account the combined propagation and reflection losses.

A cross-correlation between the bi-polar version of the transmitted sequence and the received samples yields c(m) with

$$c(m) = \sum_{a=0}^{N-1} d_{BP}(a) \cdot s(a+m). \quad (7)$$

Placing the sampled signal $s(k)$ from (6) into (7), the cross-correlation yields

$$c(m) = r \sum_{a=0}^{N-1} \sum_{n=0}^{N-1} d_{BP}(a) \cdot d_{TX}(n) \cdot p[t_0 - t_p + (m+a-n)T] \quad (8)$$

After defining $b = n - a$, using (2), and reordering the summations, we obtain

$$c(m) = r \sum_{b=-\infty}^{\infty} \left\{ p[t_0 - t_p + (m-b)T] \cdot \sum_{a=0}^{N-1} d_{BP}(a) \cdot d_{TX}(a+b) \right\} \quad (9)$$

Using (3)-(5), the inner sum equals to $\frac{N}{2}\delta(b)$, such that (9) can be simplified to

$$c(m) = r \cdot \frac{N}{2} \cdot p(t_0 - t_p + mT) \quad (10)$$

Eq. (10) shows that the cross-correlation function resembles the shape of the transmitted pulses, shifted by the propagation time $t_p$. Especially, it is maximum at a sample at the propagation time, if the pulse shape is maximum at τ=0. As $t_0$ determines the relative sampling phase, by oversampling of the received signal and down-sampling with different $t_0$ the correlation samples can be fitted to the expected pulse shape, and the center of the pulse can be obtained with a high accuracy. By determining the center of the pulse, a higher accuracy can be achieved than predicted by the bandwidth of the test signal.

Fig. 2 shows an example of the measured pulse samples after 2 km and the corresponding fitted Gaussian pulse according to the equation

$$y = A_{offset} + A \cdot exp\left(\frac{-(x-\mu)^2}{\sigma}\right) \quad (11)$$

Seven sampling points were used to obtain the parameters $A$, $\mu$, $c$, $\sigma$ for a minimum quadratic error. The resulting parameter $\mu$, corresponding to the pulse delay, was used as a measure for the round-trip time to the reflection.

Results obtained with the Gaussian pulse shape were compared to other pulse shapes (e.g. raised cosine). For the propagation delay of the fiber under test, as the time difference between the reflections from the fiber input and the fiber end, there was no noticeable difference between the different pulse shapes.

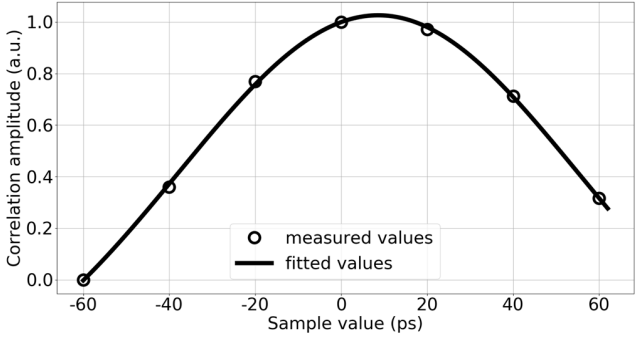

Fig. 2. Fitting of Gaussian pulse to cross-correlation function.

It should be noted that multiple reflections in the fiber, spaced by less than the sequence length, lead to overlapping reflected signals at the receiver. However, due to the linear superposition of the reflections and the cross-correlation function, both reflections can be separated after the cross-correlation, as long as the reflections are separated by more than the pulse width. For a data rate of 10 Gbit/s, this spatial resolution corresponds to approximately 10 mm. To obtain a precision better than one sample period several fitting algorithms were implemented. The fitting algorithms were investigated in terms of computation time and square error of the fit. The first algorithm which was implemented was Caruana's method [9]. The coefficients of the Gaussian function in Eq. (11) were estimated with a Jacobi matrix and a residual vector. The algorithm is fast in computation time. However, the resulting square error is higher compared to the other algorithms. The second algorithm was a gradient descent algorithm, in which all four coefficients of the Gaussian function were numerically optimized. The computation time is higher than with Caruana's method, but the resulting square error is lower. The third algorithm is a combination of the two algorithms, as shown in Table 1. The coefficients $\mu$ and $\sigma$ in Eq. (11) were numerically optimized, while the coefficients $A$ and $A_{offset}$ were calculated with a Jacobi matrix in each step. This last algorithm has a faster computation time than the gradient descent algorithm but is slower than Caruana's method. This is a consequence of the number of iterative steps which are needed to minimize the square error. However, the last algorithm achieved the lowest square error of the three algorithms. Hence, we used this algorithm in our further work.

TABLE 1
COMPARISON OF FITTING ALGORITHMS

|  | Caruana's method | Gradient method | Hybrid method |
|---|---|---|---|
| Computation time (ms) | 1 | 1176 | 89 |
| Square error | $1.18 \cdot 10^{-3}$ | $5.87 \cdot 10^{-4}$ | $3.65 \cdot 10^{-5}$ |

Results obtained over 2.2 km fiber spool.

### C. Bit Sequence

For some measurements reported here, a pseudo-random bit sequence (PRBS) of length $2^7-1$ was used as the probe sequence. This sequence has a zero auto-correlation function, if the sequence is repeated in a cyclic way. However, the aperiodic auto-correlation function of the non-periodic signal shows pre- and post-cursors. To suppress these outlying correlation peaks, a filter function was applied to the correlation function in the time domain. Most results, however, were obtained with transmitting two complementary Golay sequences [10] in consecutive measurements and adding the correlation functions. Golay sequences are well suited to suppress pre- and post-cursors, as these are inverted for complementary functions and therefore cancel out upon addition such that only the main correlation peak remains. To compare different code lengths, the signal to noise ratio (SNR) for the reflection peak after 2.2 km was calculated as the ratio of the correlation peak and the root mean square of 1000 surrounding noise values. The SNR over the code length is shown in Fig. 3 for various number of averaged traces. It can be seen that an SNR increase of approximately 3 dB for each doubling of the code length was achieved, which is consistent with the corresponding increase of the correlation peak. When increasing the number of averages, initially an SNR gain was achieved as shown in Fig. 4. For more than approximately 500 averages, the SNR was not further improved.

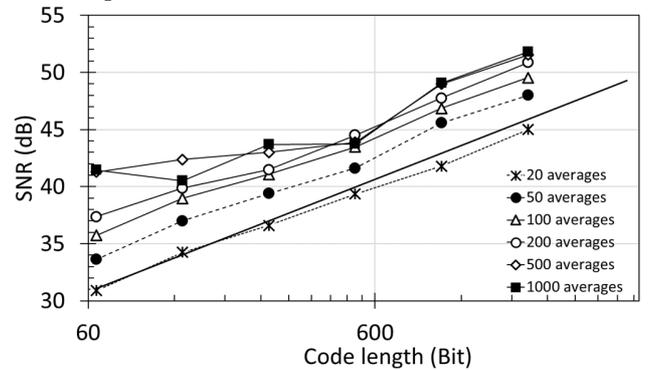

Fig. 3. SNR as a function of code length. The solid line without markers corresponds to a slope of 3 dB per doubling of the code length.

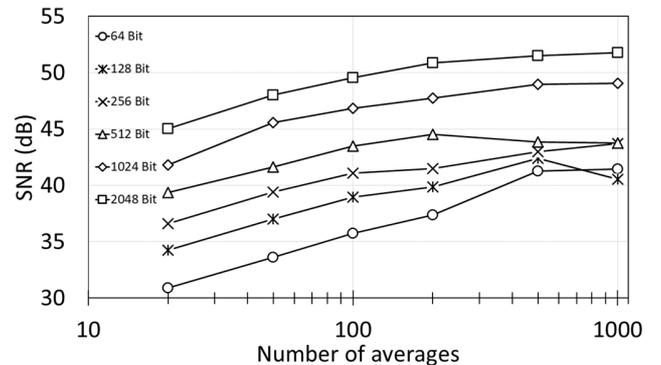

Fig. 4. SNR as a function of the number of averaged signal traces.

## III. Measurement Accuracy

The accuracy of the measurement depends on several parameters. First, the wavelength accuracy of the probe signal, in combination with the fiber chromatic dispersion (CD), can lead to a measurement error. Assuming a CD of 16 ps/nm/km, as for standard fiber in the C-band, and a laser wavelength accuracy of ±20 pm (±2.5 GHz), the round-trip delay error would grow linearly with length with 0.32 ps/km.

Secondly, the clock accuracy, with which the reflected signals are recorded, directly impacts the time delay accuracy. A clock accuracy of ±1 ppm corresponds to a round-trip time error of 5 ps/km, growing linearly with the fiber length. Thirdly, the accuracy to determine the reflection point, is impacted by the measurement noise and the pulse shape. For a long fiber, CD might impact the measurement. However, while CD broadens the pulses, the center of the pulses is not shifted, as long as the spectrum is symmetric, and the pulses are unchirped. In the following, the latter two impacts are further discussed. A fourth parameter impacting the round-trip delay measurement is the temperature of the fiber under test. While this parameter impacts the refractive index and the length of the fiber, and therefore the signal delay, this parameter actually changes the value to be measured and needs to be considered, when consecutive measurements are taken to determine the repeatability of the measurement.

### A. Impact of Clock Accuracy

One important factor for the accuracy of the propagation delay is the accuracy of the clock used as time base during the measurement, as this linearly translates into the accuracy of the derived delay. The internal clock of the real-time oscilloscope (DPO77002SX) was specified with an accuracy of ±0.8 ppm within the first year [11], which would translate, for the delay of 100 km fiber (approx. 500 µs), to an accuracy of ±400 ps. To improve the timing accuracy, we synchronized the real time oscilloscope with an external oven-controlled crystal oscillator (OCXO) with a stability of ±5 ppb [12]. During the experiment, the OCXO was always synchronized by around five GNSS satellites to guarantee the long-term time accuracy. To measure the accuracy improvement from using the external clock, we recorded the external OCXO-based 10-MHz synchronization signal on the oscilloscope with its internal clock as time base. A trace of 100,000 clock periods were recorded, and the duration was measured to be 9.9999756 ms, which is a deviation of -24.4 ns from the expected 10 ms. This results in an error of the internal clock frequency of +2.44 ppm. The external clock accuracy of ±5 ppb would mean an accuracy of ±2.5 ps for the delay of 100 km fiber.

However, if only differential delay is of interest and two fibers can be characterized simultaneously with the same clock, a much larger clock offset can be accepted. For instance, if a delay asymmetry of 10 ns is measured between two fibers, even a clock accuracy of ±100 ppm, as achievable with standard telecom oscillators, yields an error of only 1 ps on this delay asymmetry.

### B. Measurement of Reflection Accuracy and Repeatability

The accuracy of the determination of the reflection point can be estimated in two ways. First, we considered how cascading the propagation length results in a proportional increase in propagation time. Second, we measured the propagation delay in multiple consecutive tests. For the cascading approach, the fiber under test was a 2.2 km fiber spool with high reflections at the input (~30%) and the output (~100%). The transmitted data sequence was then observed, first, after a direct reflection from the fiber input, second, after a single round-trip over the fiber length and reflection at the fiber end, and, third, after a double round-trip with reflections at the fiber end, fiber input, and, again, at the fiber end. The time difference between the first and the second received signals corresponds to the round-trip propagation time between the input and output reflectors. Likewise, the time difference between the second and third received signal corresponds to the double round-trip propagation. For a round-trip time of approximately 21.6 µs, we measured a difference of 1.9 ps between the first and double round-trip time. This measurement was performed with a 127-bit PRBS sequence at a data rate of 10 Gbit/s and 1000 trace averages on the oscilloscope. In Fig. 5, the averaged traces are shown. In the inset of the figure, the received reflected PRBS traces from the fiber input and after a double round-trip are shown. In Fig. 6, the correlation function of this measurement is shown, demonstrating the three correlation peaks after 0, 22 and 44 µs. The insets show the obtained correlation peaks in more detail. Here, the reflection peak pulse width of approximately 100 ps is noticeable. The SNR of this measurement series was 16.9 dB, 17 dB, and 5.8 dB of the reference reflection, round-trip time reflection and double round-trip reflection, respectively.

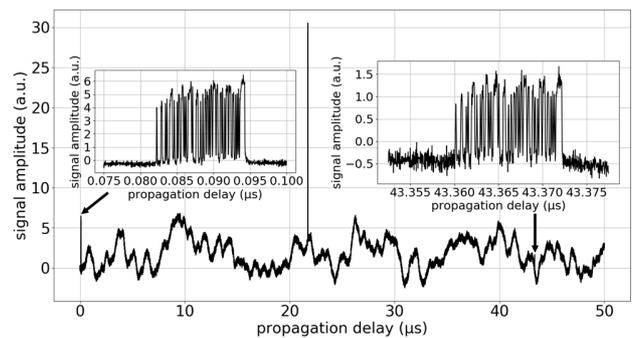

Fig. 5. Received signal trace from a 2.2-km fiber before correlation. In the inset, the reflections from the reference and after double round-trip are shown. The burst width of both is 12.7 ns.

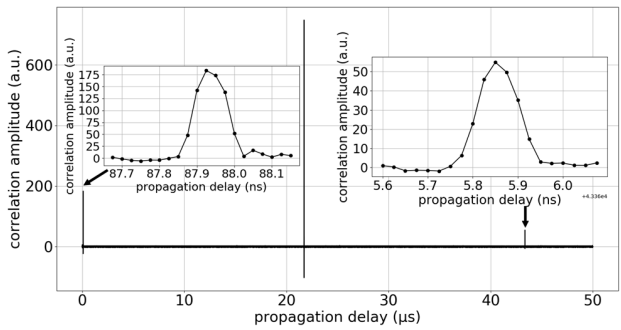

Fig. 6. Cross-Correlation of received signal from a 2.2- km fiber with the transmitted PRBS sequence. The insets show the correlation peaks from the reference and the double round-trip signal, each with a pulse width of approximately 100 ps.

To determine the impact of the trace averaging, sub-sets of the 1000 measurements were taken, averaged, and the round-trip times were determined. The RMS error as a function of the number of averages is shown in Fig. 7. For more than 150 averages, the RMS difference between first and second round trip, representing the measurement error, is below 3 ps.

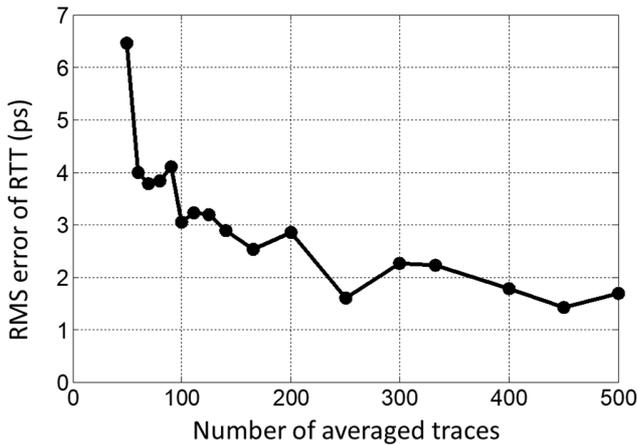

Fig. 7. RMS difference between first and second round-trip time over 2.2 km fiber as function of average traces.

The repeatability of the measurement was determined by repeating the fiber delay measurement several times. Fig. 8 shows the round-trip evolution for a 100 km fiber. The values are offset by 990.537 µs, which was the delay measured in the first test. In this experiment, two complementary Golay sequences at a data rate of 2 Gbit/s were used as probe signal and averaged over 4000 samples. During the 3-hour measurement, the measured latency increased by approximately 1200 ps, or 1.2 ppm, which can be attributed to a temperature increase of approximately 0.2 K in the laboratory. To reduce the impact of the temperature change on the accuracy estimation, a 4th order curve over time was fitted to the measured delay values, assuming a smooth increase in fiber temperature over time. A 12 ps RMS deviation of the measured values from this curve was calculated for the 100 km fiber round-trip time [13].

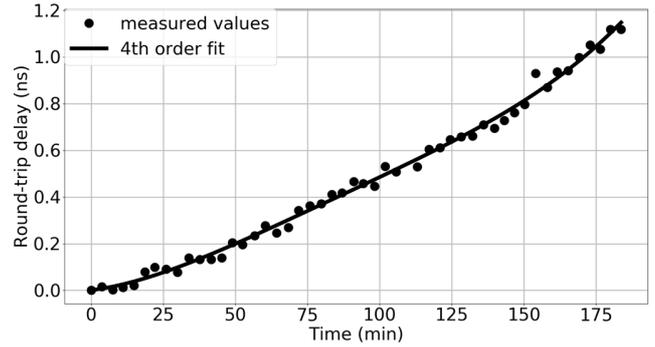

Fig. 8. Measured round-trip delay of 100 km fiber over 3 hours; values offset by 990.537 µs.

## IV. MONITORING OF LINK DELAY

### A. Setup for Seamless Delay Measurement

To monitor the delay of a full transmission link, all sections, the transmission fiber as well as optical components and sub-systems, need to be taken into account [14]. For an accurate delay measurement, precise demarcations between the sections are required. These demarcation points can be realized as reflectors, as shown in Fig. 9. When the link section allows the probe signal to propagate in both directions, like the fiber between nodes, reflectors at both sides of the section can be used to determine the round-trip delay of the probe signal for the respective section. If only uni-directional propagation of the probe signal is possible, e.g. because an isolator is contained in the section, as in an EDFA, the propagation delay can be obtained by measuring the single-pass delay through the section and correcting this value for the measurement delay to and from the demarcation points.

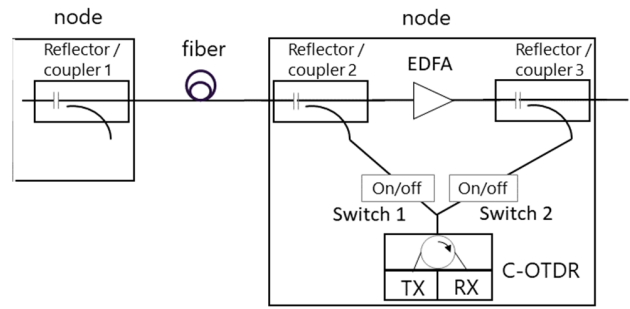

Fig. 9. Measurement setup for monitoring of link delay with fiber and node sections.

The single-pass measurement can use the same equipment as the reflective C-OTDR measurement by sending the propagated signal from the output of the section directly to the receiver. To maintain synchronization between transmitter and receiver, though, this requires both to be collocated, such that only sections contained within one network node location can be measured.

A measurement procedure for the whole link, could consist of the following three steps, as explained in reference to Fig. 9:

1. Turn on only Switch 1 and transmit the probe signal. Receive the signal from the near end reflector 2 and from the far end reflector 1 and calculate the round-trip fiber delay of the attached fiber span as the difference between both. Also record the round-trip delay to reflector 2.

2. Turn on only Switch 2 and transmit the probe signal. Receive the reflected signal from reflector 3. Record the round-trip delay to reflector 3.

3. Turn on Switch 1 and Switch 2 and transmit the probe signal. Receive the signal through the uni-directional device and calculate the single-pass delay between reflectors 2 and 3 by subtracting half the round-trip delays to these reflectors, as recorded in steps 1 and 2, from the measured single-pass delay.

For components, which cannot be measured with the probe signal, because they are wavelength-specific or because they need to be measured during operation, minimum and maximum delay specifications might be used to get a lower and upper limit on the propagation delay. Rec. ITU-T G.671 [15] has recently added the group delay as a parameter for optical components and sub-systems.

### B. Monitoring the Delay of Long Fiber

For the single-ended measurement of the delay of a long fiber between the nodes, the probe signal must travel twice along the fiber, in forward and backward direction. For the results shown above in Fig. 8, a data rate of 2 Gbit/s and an electrical receiver filter with a 2-GHz bandwidth were used to reduce the effects of chromatic dispersion of the 100 km standard single mode fiber (SSMF) and the noise accumulation. To increase the precision, a higher data rate with narrower data pulses would be preferable. At a data rate of 10 Gbit/s, however, chromatic dispersion accumulated along 200 km SSMF would strongly broaden the signal pulses. The introduction of chromatic dispersion compensation components would mitigate the pulse broadening at 10 Gbit/s.

Therefore, we introduced two chirped fiber Bragg gratings with a dispersion of -1700 ps/nm per grating, one each between TX and circulators and between circulator and RX. The broadening of the pulse reflected at the fiber end was compensated, while the reference pulse was broadened due to the negative dispersion. As the latter is received with a higher signal-to-noise ratio, the broadening is acceptable, as long as the center of the pulse can still be determined with sufficient accuracy. To evaluate the measurement accuracy, the group delay for a 100-km fiber span was measured by the C-OTDR method in 10-minute intervals. Time-interleaved, single-pass measurement were performed with the same equipment. The results are shown in Fig. 10, where the single-pass delay and half of the round-trip time was offset by 495.250 µs. During the measurement time, the temperature fluctuated slowly. A fourth order polynomial function was fitted to each set of the measured group delay values, and an average offset of 3.9 ps between both curves was observed [16]. This constant difference, corresponding to approximately 800 µm fiber length, is probably due to a mis-calibration of the single-pass measurement.

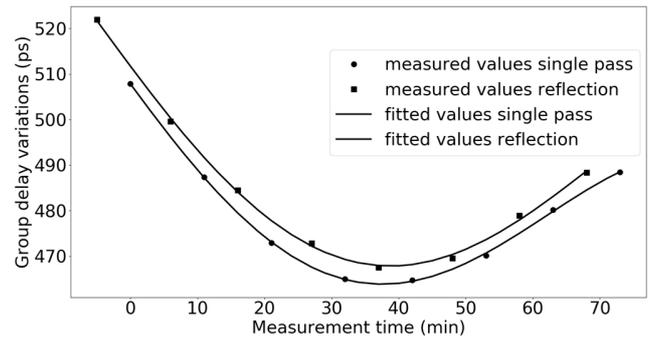

Fig. 10. Comparison of reflection measurement and single pass measurement over 70 minutes with 4th order polynomial fits to both series

### V. TEMPERATURE DEPENDENCE OF FIBER DELAY

#### A. Short Jumper Cables

With respect to some of the measurements presented above, it was already noted that temperature variations of the fiber lead to a change in the propagation delay time. A typical thermal coefficient of delay (TCD) is approximately 7 ppm/K for bare fiber [17][16]. Most of this change is due to the temperature coefficient of the refractive index. It is also known, however, that physical strain on the fiber changes the refractive index and the propagation delay. For fibers with a tight-buffer inner jacket, with different expansion coefficients between the materials, temperature changes can also lead to the introduction of strain and therefore to a variation in the TCD. In [18], the delay of a 25-m jumper cable was measured for a temperature range from 10 to 70°C using the C-OTDR technique. As shown in Fig. 11, the delay changed non-linearly over temperature with a larger TCD on the order of 40 ppm/K for low temperatures, approaching 7 ppm/K for temperatures above 50°C. The measurement was performed with increasing and decreasing temperatures, shown in hollow and solid markers, respectively. The temperature settling time between consecutive measurements was 20 minutes. For the cable with both connectors attached, as indicated by the solid lines, an inverse hysteresis can be observed. In the cooling cycle, the delay decreased faster, while in the heating cycle the delay increase was faster. This can be taken as an indication that the temperature change introduced strain, which was partly settled during the measurements at higher temperatures and then inverted, when the fiber was cooled. This behavior was not observed, after one of the connectors of the jumper cable was cut off, as shown by the dashed lines, where both measurement results overlap. This is also an indication that the inverse hysteresis is induced by the stress due to the connector attachment.

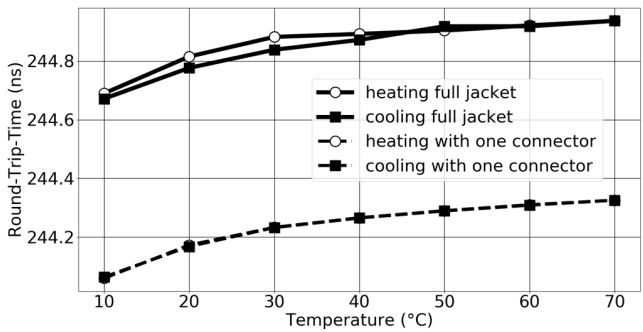

Fig. 11. Temperature dependence of delay of 25-m jumper cable with outer jacket and connectors (solid lines) and with one connector cleaved off (dashed lines), for in-creasing (hollow markers) and decreasing (solid markers) temperatures.

### B. Fiber in Buried Cable

The delay variations of a buried fiber cable were monitored over a longer period using the C-OTDR technique. The delay of four 8.5 km long cabled and deployed fibers was continuously monitored over a period of two weeks during the summer 2018 [19]. While the outside temperature varied by 15 K during this time, as recorded by a nearby weather station [20], the round-trip fiber delay variations were measured to be 800 ps. The delay differences between the fibers were up to 12 ps during this time. It was observed that the delay changes only slowly followed the outside temperature variations. We assumed the buried fiber temperature to follow the outside temperature with a first-order low-pass filter function and extracted a time constant of 12.7 days for this filter. The TCD of the cable, also extracted during this fitting process, was 7.5 ppm/K.

The extracted time constant was applied to the weather station temperature data to calculate the expected temperature evolution of the cabled fiber over a year. Even though the fiber was buried about one meter deep, it experienced a temperature swing of more than 25 K, as shown in Fig. 12. This value was verified by ground temperature data from Potsdam Institute for Climate Impact Research for a 1-m depth in a similar region in Germany [21]. Also shown as dots in Fig. 12, is the varying delay of one of the cabled fibers, offset by 41.038 μs, (fiber #2), measured every few months over the duration of a year. While for the higher temperatures the TCD is around 8 ppm/K, for the transition to temperatures around 0 °C it is about twice as large. The right-hand delay scale compares to the left-hand temperature scale with a TCD of 14.6 ppm/K.

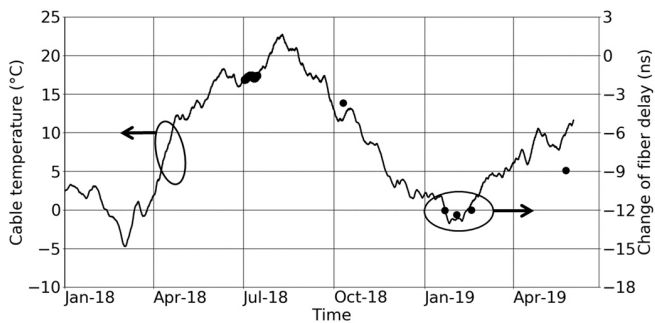

Fig. 12. Calculated long-term variation of buried fiber temperature (line, left hand axis) and measured fiber delay variation (dots, right hand axis) over time.

Between the fibers in the same cable, a delay difference of 10 ns between the "slowest" and the "fastest" fiber was observed. These delay differences were not constant but also changed with temperature. Fig. 13 shows the temperature variation of the delay differences of fibers #1, #3, and #4 with respect to the fastest fiber #2, offset by the value measured at 17 °C. It can be seen that the differential delay changes by up to 100 ps over 18 K for this 8.5-km span of standard single mode fiber. It can be expected that, for longer spans, the delay difference increases approximately linearly with length such that for an 80-km span the latency difference between different fibers of a cable can change by as much as a nanosecond over the year. This raises the requirement to monitor the fiber delay during operation for synchronization applications, even if initial delay differences are measured and calibrated upon installation.

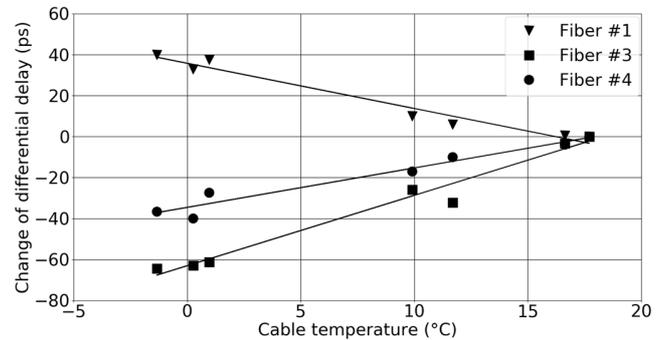

Fig. 13. Variation with cable temperature of delay differences between fibers in a cable. Lines are linear trends.

### VI. MEASUREMENT OF FIBER CHROMATIC DISPERSION

As the propagation delay over the fiber can be measured from one fiber end, this method can also be used to determine the chromatic dispersion of a deployed fiber with access to a single fiber end, potentially during operation of the system. For this purpose, the fiber propagation delay is measured at different wavelengths, and the derivative of the propagation delay with respect to wavelength determines the net chromatic dispersion. If the refractive index of the fiber is known, which is typically the case with an accuracy of ~0.1%, the absolute value of the propagation delay can be used to determine the length of the fiber section to yield the chromatic dispersion coefficient. When the different wavelength points are measured over a longer period of time (more than a few minutes), care must be taken to consider the temperature changes of the fiber, which lead to a change of the propagation delay and to an error in the chromatic dispersion measurement. For a 2.2-km fiber, the propagation delay was measured for 7 wavelengths in the range from 1530 to 1566 nm in a non-monotonic order. For each of the wavelength points, 1000 traces were recorded and saved for further processing. This series of measurements took approximately 3.5 hours to complete. The recorded data were then evaluated in blocks of 250 traces to calculate the propagation delay. Fig. 14 shows, for each of the wavelengths, the evolution of the propagation delay over the four blocks. By extrapolating the thermal expansion between blocks 3 and 4 of one wavelength to block 1 of the next measured wavelength, we calculated the delay change of the fiber over the measurement time of 3.5 hours, as shown in Fig. 15.

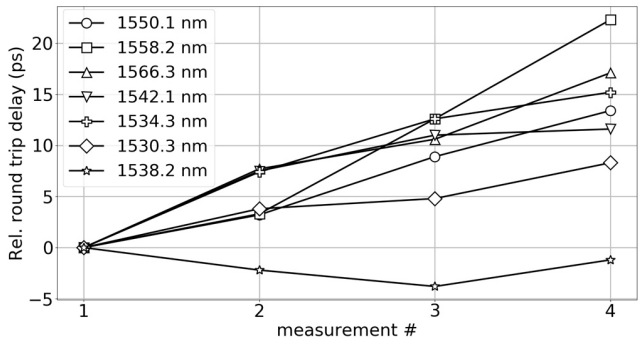

Fig. 14. Change of round-trip delay over consecutive blocks of 250 traces.

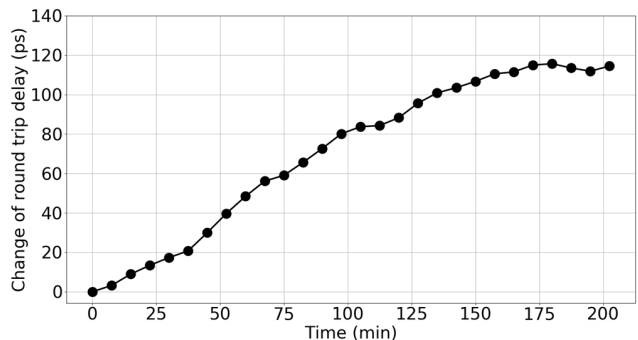

Fig. 15. Change of round-trip delay over time of fiber under test

Similar results were obtained using block lengths of 100 or 200 traces. The resulting round-trip delay increase of 120 ps (5.45 ppm of the fiber round-trip delay) corresponds to a temperature increase of approximately 0.8 K and would have resulted in a CD calculation error of 0.75 ps/(nm·km). After taking into account the delay changes over time, a 2nd order polynomial was fit to the delay vs. wavelength curve. A CD at 1550 nm of 16.698 ps/(nm·km) and a CD slope of 0.0588 ps/(nm$^2$-km) were obtained. Over the whole C-band, the measured CD values differed by a maximum of 0.04 ps/(nm·km) from the values measured using a modulation phase delay CD analyzer.

## VII. Further Applications

The highly accurate measurement of the propagation delay in fiber enables additional sensing application, which have so far been based on an evaluation in the frequency domain. Using a chirped fiber grating (CFG) with a dispersion of 2000 ps/nm, for example, a temperature sensor can be realized [22] with a delay temperature coefficient of approximately 20 ps/K, such that a C-OTDR with an accuracy of 1 ps can measure the temperature with an accuracy of 0.05 K. Multiple identical CFG sensors can be cascaded to be probed by a single C-OTDR. Based on the change of delay with temperature or strain, a simple fiber jumper could also be used as a sensor for these quantities without the need for Bragg gratings. The C-OTDR therefore appears to be a versatile technique for sensing applications.

## VIII. Implementation of C-OTDR with Telecommunication Components

To enable deployment and measurements in the field, the C-OTDR function described in the previous sections was implemented in a multiprocessor system-on-chip (MPSoC), and an SFP was used as the optical front end [8]. A field-programmable gate array (FPGA) as part of the MPSoC substituted the arbitrary waveform generator and the real-time oscilloscope, performing signal generation, recording, and processing. The offline processing of the signals was performed by a microprocessor implemented in the MPSoC. With Linux as operating system, the Python laboratory signal processing scripts could be re-used.

While the real-time oscilloscopes in the lab provided an amplitude resolution of at least 7 bits, a binary high-speed data input (MGT) was used to read the signal into the FPGA. The amplitude resolution was improved by averaging multiple signal traces, superimposed by noise from Rayleigh backscattering and thermal noise of the receiver.

Oversampling of the received signal was achieved by shifting the sampling times in the MGT receiver in steps of a fraction of the bit rate. This feature of the FPGA enabled a higher temporal resolution than defined by the data rate. To avoid timing problems, the sampling clock and the transmitter clock were phase aligned by sending a trigger signal at the beginning of each probe burst. Time delay measurements with the integrated setup showed a similar performance as with the laboratory setup, as reported in [8].

## IX. Conclusion

We presented a Correlation-OTDR as a precise method to measure propagation delay in optical fibers. The accuracy of the propagation delay is proportional to the equipment clock accuracy. A clock uncertainty of ± 1 ppm would lead to a delay error of ± 500 ps for 100 km of fiber. A higher accuracy can be achieved by synchronization of the measurement clock e.g. to a satellite time signal. The measurement accuracy is also impacted by the determination of the reflection point. We measured an accuracy of 1.9 ps for a 2.2-km fiber. A comparison of single pass and reflection measurement was used for longer fibers up to 100 km. We achieved an accuracy of 3.9 ps, using dispersion compensating gratings, for the delay of 100 km fiber. The C-OTDR method was used to demonstrate the temperature-dependent propagation delay of jumper cables and fiber in a deployed cable. It was shown that over a year the delay of buried fiber can change by more than 1 ns/km.